\documentclass[12pt]{article}

\usepackage{latexsym}
\usepackage{amssymb,amsfonts,amsmath}
\usepackage{graphicx} 
\usepackage{indentfirst}
\usepackage{bbm}
\usepackage{amssymb}
\usepackage{verbatim}
\usepackage{amsmath, amsthm,amssymb}
\usepackage{mathrsfs}
\usepackage{hyperref}
\usepackage{amsfonts}
\usepackage{dsfont}
\usepackage{cite}
\usepackage{xcolor}
\usepackage[multiple]{footmisc}

\parskip=\medskipamount

\newcommand {\cL}{{\cal L}}

\def\a{\alpha}

\def\b{\beta}

\def\d{\delta}

\def\f{\phi}
\def\g{\gamma}

\def\s{\sigma}

\newcommand{\ad}{{\dot{\alpha}}}                           
\newcommand{\bd}{{\dot{\beta}}}    
  
\newcommand{\ve}{\varepsilon}                            

\newcommand{\hf}{\frac12}

\newcommand{\vf}{\varphi}

\newcommand{\be}{\begin{equation}}
\newcommand{\ee}{\end{equation}}
\newcommand{\bea}{\begin{eqnarray}}
\newcommand{\eea}{\end{eqnarray}}

\def\double #1{#1{\hbox{\kern-2pt $#1$}}}

\newcommand{\gd}{{\dot\g}}
\newcommand{\dd}{{\dot\d}}

\newif\ifdtup

\newcommand{\bsubeq}{\begin{subequations}}
\newcommand{\esubeq}{\end{subequations}}

\numberwithin{equation}{section}

\newcommand{\sSL}{\mathsf{SL}}

\begin{document}

\begin{titlepage}
\begin{flushright}
December, 2019\\
\end{flushright}
\vspace{5mm}

\begin{center}
{\Large \bf Generalised conformal higher-spin fields in curved backgrounds}\\ 
\end{center}

\begin{center}

{\bf   
Sergei M. Kuzenko and Michael Ponds 
} \\
\vspace{5mm}

\footnotesize{
{\it Department of Physics M013, The University of Western Australia\\
35 Stirling Highway, Perth W.A. 6009, Australia}}  
\vspace{2mm}
~\\
Email: \texttt{ 
sergei.kuzenko@uwa.edu.au, michael.ponds@research.uwa.edu.au}\\
\vspace{2mm}

\end{center}

\begin{abstract}
\baselineskip=14pt
The problem of constructing gauge-invariant actions for conformal higher-spin 
fields in curved backgrounds is known to be notoriously difficult.
In this paper we present gauge-invariant models for conformal maximal depth  fields with spin $s=5/2$ and $s=3$ in four-dimensional Bach-flat backgrounds. We find that certain lower-spin fields must be introduced to ensure gauge invariance when $s>2$, which is analogous to a conjecture made earlier in the literature for conformal higher-spin fields of minimal depth. 
\end{abstract}

\vfill

\vfill
\end{titlepage}

\newpage
\renewcommand{\thefootnote}{\arabic{footnote}}
\setcounter{footnote}{0}

\tableofcontents{}
\vspace{1cm}
\bigskip\hrule

\allowdisplaybreaks


\section{Introduction}

Conformal higher spin (CHS) models in four-dimensional  Minkowski space 
\cite{FT} were proposed more than thirty years ago.\footnote{The CHS gauge fields 
introduced in \cite{FT} are naturally realised as component fields of the conformal higher spin supermultiplets  \cite{HST,KMT}.} 
A few years later, CHS superalgebras \cite{FL-algebras}
and associated gauge theories in the cubic approximation
\cite{FL-vertices} were constructed, as an extension of the seminal work by Fradkin and Vasiliev on higher-spin superalgebras \cite{FV1,FV2,Vasiliev88} and interacting massless higher-spin theories \cite{FV-vertices}.
Finally, the Lagrangian formulation for a complete interacting bosonic CHS theory 
was sketched \cite{Tseytlin} and fully developed \cite{Segal} in 2002; see
also \cite{BJM1,BJM2,Bonezzi} for more recent related studies.
However,  gravitational interactions of CHS fields still remain quite mysterious.

For every positive integer or half-integer $s \geq 1$, the gauge-invariant action for a conformal spin-$s$ field contains $2s$ derivatives, and therefore it is a higher-derivative theory for $s>1$.
This higher-derivative structure implies that  
the problem of a consistent deformation of CHS actions
from flat to curved gravitational backgrounds is nontrivial.
For instance, since the 1985 work by Fradkin and Tseytlin \cite{FT}  
it was reasonably clear
that there should exist a consistent formulation for all CHS models on arbitrary conformally flat backgrounds. However, such a formulation has been developed only recently \cite{KP2}, and it also works for generalised CHS models. The latter describe conformal spin-$s$ fields of depth $t>1$, which  have been studied for more than thirty years 
 \cite{DeserN, DeserN2, Vasiliev2006,Vasiliev2009,DNW,DeserW5,DeserW6,BG2013,GrigorievH, Skvortsov}.
Dynamics of the conformal graviton ($s=2$) can be consistently defined on an arbitrary Bach-flat background,
since the corresponding gauge-invariant model is obtained by linearising the Weyl gravity action about its stationary point, and the equation of motion for conformal gravity is that the Bach tensor vanishes.   
The same is true of the conformal gravitino ($s=3/2$), since the corresponding gauge-invariant model is obtained by linearising the action for conformal supergravity \cite{KTvN1}. 

For quite some time it was believed that the dynamics of a single conformal  spin-$s$ field could be consistently defined on Bach-flat backgrounds 
for any $s>2$, see e.g. \cite{NT,GrigorievT}.
However, recent studies of the conformal spin-3 theory
\cite{NT,GrigorievT,BeccariaT,Manvelyan} have demonstrated \cite{GrigorievT,BeccariaT} that 
 gauge invariance of a pure spin-3 field can only be upheld to first order in the background curvature. It was then conjectured by Grigoriev and Tseytlin \cite{GrigorievT} that it might be possible to restore gauge invariance by switching on a coupling to a conformal spin-1 field. 
 This idea has been confirmed  by Beccaria and Tseytlin \cite{BeccariaT} who 
explicitly worked out the spin 1--3 mixing terms.
 Nevertheless, due to its higher-derivative nature, the pure spin-3 sector to all orders in the background curvature is still unknown, and the story of the conformal spin-3 field 
 in curved backgrounds remains so far unfinished. 
 
 It appears that new insights into the problem under consideration may be obtained by studying somewhat simpler dynamical systems -- generalised CHS fields
   in a gravitational background. The point is that one can decrease the number of derivatives appearing in the action by increasing the depth of the gauge transformations. This bypasses some of the technical difficulties associated with higher-derivative models
 such as the conformal spin-3 one. It is for this reason that in this paper we concentrate on
 conformal maximal depth (CMD)  fields and work out the cases  $s=5/2$ and $s=3$.
 
 This paper is organised as follows. In section \ref{section2} we summarise the basics of generalised conformal fields and review their gauge invariant formulations on arbitrary conformally flat backgrounds. Section \ref{section3} reviews the extension of the CMD spin $s=2$ model to Bach-flat backgrounds. In sections \ref{section4} and \ref{section5} we demonstrate how one can make use of lower-spin fields to achieve a gauge invariant description of CMD spin $s=5/2$  and $s=3$ fields in a Bach-flat background, respectively. Concluding comments are given in section \ref{section6}.
The main body of the paper is accompanied by a technical appendix.


\section{Generalised conformal gauge fields}\label{section2}

Throughout this work we make use of the conformal calculus described in \cite{KP2} 
(building on the earlier work \cite{BKNT-M1}), to where we refer the reader for further details. The parts of this formalism that are essential to the models constructed in this paper are as follows. 

In modern approaches to conformal gravity \cite{KTvN1}, the structure group of the space-time manifold is promoted from the Lorentz group to the conformal group. 
The geometry of space-time is then described by the conformally covariant derivative
\begin{align}
\nabla_{a}=e_{a}{}^{m}\partial_m-\frac{1}{2}\omega_{a}{}^{bc}M_{bc}-\mathfrak{b}_a\mathbb{D}-\mathfrak{f}_{a}{}^{b}K_b \label{1.1}
\end{align}
 where $M_{bc}, \mathbb{D}$ and $K_a$ are the Lorentz, dilatation and special conformal generators respectively. Upon imposing appropriate constraints on the torsion and curvature
 tensors, one can show that 
 the algebra of conformal covariant derivatives in the two-component spinor notation  (we adopt the spinor conventions of \cite{BK})
 takes the form
\bea \label{1.2}
\big[\nabla_{\a\ad},\nabla_{\b\bd} \big]&=&-\big(\ve_{\ad\bd}C_{\a\b\g\d}M^{\g\d}+\ve_{\a\b}\bar{C}_{\ad\bd\gd\dd}\bar{M}^{\gd\dd}\big) \notag\\
&&
-\frac{1}{4}\big(\ve_{\ad\bd}\nabla^{\d\gd}C_{\a\b\d}{}^{\g}+\ve_{\a\b}\nabla^{\g\dd}\bar{C}_{\ad\bd\dd}{}^{\gd}\big)K_{\g\gd}~.
\eea
Here $C_{\a\b\g\d}$ and $\bar C_{\ad\bd\gd\dd}$ are the self-dual and anti self-dual parts of the Weyl tensor and are related to the Weyl tensor $C_{abcd}$ through
\begin{subequations} \label{666}
\begin{align}
C_{\a(4)}&= \frac{1}{2}(\s^{ab})_{\a(2)}(\s^{cd})_{\a(2)}C_{abcd}~,\\
\bar{C}_{\ad(4)}&= \frac{1}{2}(\tilde{\s}^{ab})_{\ad(2)}(\tilde{\s}^{cd})_{\ad(2)}C_{abcd}~,\\
C_{abcd}&=\frac{1}{2}(\s_{ab})^{\a(2)}(\s_{cd})^{\a(2)}C_{\a(4)}+\frac{1}{2}(\tilde{\s}_{ab})^{\ad(2)}(\tilde{\s}_{cd})^{\ad(2)}\bar{C}_{\ad(4)}~.
\end{align}
\end{subequations}
The commutation relations 
\eqref{1.2} should be accompanied by 
\bea
\big[\mathbb{D},\nabla_{\b\bd} \big]=\nabla_{\b\bd}~,\qquad \big[K_{\a\ad},\nabla_{\b\bd}\big] &=& 4\ve_{\ad\bd}M_{\a\b}+4\ve_{\a\b}\bar{M}_{\ad\bd}-4\ve_{\a\b}\ve_{\ad\bd}\mathbb{D}~.
\eea

Models for generalised conformal higher-spin gauge fields \cite{DeserN, DeserN2, Vasiliev2006,Vasiliev2009,DNW,DeserW5,DeserW6,BG2013,GrigorievH, Skvortsov} in arbitrary conformally flat backgrounds were first constructed in \cite{KP2}. Below we summarise their main properties.

Given a conformal gravity background, 
a generalised conformal gauge field $\phi^{(t)}_{\a(m)\ad(n)}$ is characterised by three positive integers $m,n$ and $t$. The first two specify the Lorentz type of  
$\phi^{(t)}_{\a(m)\ad(n)}$. This field transforms in the representation $(m/2, n/2)$ of 
$\sSL(2, {\mathbb C})$ and
 is usually said to carry spin $s=\frac 12 (m+n)$. The third integer $t$ is known as the depth, and it determines the number of derivatives that appear in the gauge transformation
 of $\phi^{(t)}_{\a(m)\ad(n)}$,
\begin{align}
\delta_{\lambda}\phi^{(t)}_{\a(m)\ad(n)}=\nabla_{(\a_1(\ad_1}\cdots\nabla_{\a_t\ad_t}\lambda^{(t)}_{\a_{t+1}\dots\a_m)\ad_{t+1}\dots\ad_n)}~,\qquad 1\leq t \leq \text{min}(m,n)~. \label{gt}
\end{align}
The ordinary CHS fields with $|m-n|\leq 1$ and $t=1$ are sometimes referred to as 
Fradkin-Tseytlin fields \cite{BG2013}.

 In order for the gauge field $\phi^{(t)}_{\a(m)\ad(n)}$ and the gauge parameter 
 $\lambda^{(t)}_{\a (m-t) \ad (n-t) }$ in \eqref{gt} to 
 be primary (i.e. annihilated by $K_{\a\ad}$), the field must have 
 the following (Weyl) weight
\begin{align}
\mathbb{D}\phi^{(t)}_{\a(m)\ad(n)}=\Big(t+1-\frac{1}{2}(m+n)\Big)\phi^{(t)}_{\a(m)\ad(n)}~. \label{weight}
\end{align}
From $\phi^{(t)}_{\a(m)\ad(n)}$ we may construct generalised higher-spin Weyl tensors \cite{KP2}
\begin{subequations}\label{weyl}
\begin{align}
\hat{\mathfrak{C}}^{(t)}_{\a(m+n-t+1)\ad(t-1)}(\phi)&= \nabla_{(\a_1}{}^{\bd_1}\cdots\nabla_{\a_{n-t+1}}{}^{\bd_{n-t+1}}\phi^{(t)}_{\a_{n-t+2}\dots\a_{m+n-t+1})\bd_1\dots\bd_{n-t+1}\ad_1\dots\ad_{t-1}}~, \label{weyl1}\\
\check{\mathfrak{C}}^{(t)}_{\a(m+n-t+1)\ad(t-1)}(\bar{\phi})&=\nabla_{(\a_1}{}^{\bd_1}\cdots\nabla_{\a_{m-t+1}}{}^{\bd_{m-t+1}}\bar{\phi}^{(t)}_{\a_{m-t+2}\dots\a_{m+n-t+1})\bd_1\dots\bd_{m-t+1}\ad_1\dots\ad_{t-1}}~,\label{weyl2}
\end{align}
\end{subequations}
 which are primary and possess the Weyl weights
\begin{subequations}
\begin{align}
&\mathbb{D}\hat{\mathfrak{C}}^{(t)}_{\a(m+n-t+1)\ad(t-1)}(\phi)=\Big(2-\frac{1}{2}(m-n)\Big)\hat{\mathfrak{C}}^{(t)}_{\a(m+n-t+1)\ad(t-1)}(\phi)~,\\
&\mathbb{D}\check{\mathfrak{C}}^{(t)}_{\a(m+n-t+1)\ad(t-1)}(\bar\phi)=\Big(2-\frac{1}{2}(n-m)\Big)\check{\mathfrak{C}}^{(t)}_{\a(m+n-t+1)\ad(t-1)}(\bar\phi)~.
\end{align}
\end{subequations}
 
Let $\cL$ be a primary scalar field of weight $+4$. Associated with $\cL$ is the functional 
\bea
S=\int \text{d}^4x \, e \, \cL~, \qquad e^{-1}=\det (e_a{}^m)  \label{58.0}
\eea
which is invariant under the gauge group of conformal gravity. Upon degauging (see section \ref{section3}), these properties mean that the action  \eqref{58.0} is invariant under Weyl transformations. In this paper we will refer to such action functionals as primary.

In any conformally flat background, the commutator of conformal covariant derivatives \eqref{1.2} vanishes. This significantly simplifies the construction of CHS models. Indeed, in such backgrounds one can show that the generalised higher-spin Weyl tensors are gauge invariant, 
\begin{align}
C_{abcd}=0\quad\implies\quad \delta_{\lambda}\hat{\mathfrak{C}}^{(t)}_{\a(m+n-t+1)\ad(t-1)}(\phi)=\delta_{\lambda}\check{\mathfrak{C}}^{(t)}_{\a(m+n-t+1)\ad(t-1)}(\bar\phi) =0~.
\end{align}
However this is not true in a general curved background, where the gauge variation is proportional to the Weyl tensor. All of these properties mean that the associated action,
\begin{align}
S_{\text{Skeleton}}^{(m,n,t)}[\phi,\bar{\phi}]={\rm i}^{m+n}\int \text{d}^4x \, e \,\hat{\mathfrak{C}}_{(t)}^{\a(m+n-t+1)\ad(t-1)}(\phi)\check{\mathfrak{C}}^{(t)}_{\a(m+n-t+1)\ad(t-1)}(\bar{\phi}) +\text{c.c.}~, \label{2.7}
\end{align}
is primary in any background, but gauge invariant only in conformally flat ones.


\section{Conformal spin-2 model in Bach-flat background }\label{section3}

Given a gauge field $\phi^{(t)}_{\a(m)\ad(n)}$ it is clear that we may decrease the number of derivatives appearing in the action \eqref{2.7} at the cost of increasing the depth $t$ of the gauge transformations. This significantly reduces the amount of work required to perform calculations in backgrounds more general than conformally flat ones. Therefore, the remainder of this work will focus on extending the gauge invariance of the skeleton action \eqref{2.7} to arbitrary Bach-flat backgrounds for CMD fields with spin $s=2,5/2,3$. 

We begin by reviewing the gauge-invariant model for the maximal depth spin-2 field (corresponding to $m=n=t=2$) in a Bach-flat background \cite{KP2}. 
Upon degauging, its action coincides with the one studied earlier
in Ref. \cite{BT2015}, where it was suggested that gauge invariance could only be upheld in an Einstein space. 
See below for further discussion on this.

For bosonic spin-$s$ fields with $m=n=s$, we may choose the gauge field to be real,
\begin{align}
h^{(t)}_{\a(s)\ad(s)}:=\phi^{(t)}_{\a(s)\ad(s)}=\bar{h}^{(t)}_{\a(s)\ad(s)}~.
\end{align}
 This means that \eqref{weyl1} and \eqref{weyl2} coincide,
 \begin{align}
 \mathfrak{C}^{(t)}_{\a(2s-t+1)\ad(t-1)}(h):=\hat{\mathfrak{C}}^{(t)}_{\a(2s-t+1)\ad(t-1)}(h)=\check{\mathfrak{C}}^{(t)}_{\a(2s-t+1)\ad(t-1)}(h)~.
 \end{align}
 Since we will be dealing exclusively with CMD fields, we will usually drop all labels that refer to $t$ when it's value is clear from the context.

 The spin-2 field  $h_{\a(2)\ad(2)}$ is defined modulo the depth 2 gauge transformations
\begin{align} 
\delta_{\lambda}h_{\a(2)\ad(2)}=\nabla_{(\a_1(\ad_1}\nabla_{\a_2)\ad_2)}\lambda~. \label{2.10} 
\end{align}
Here both $h_{\a(2)\ad(2)}$ and $\lambda$ are primary and have Weyl weights
\begin{align}
\mathbb{D}h_{\a(2)\ad(2)}=h_{\a(2)\ad(2)}~,\qquad  \mathbb{D}\lambda=-\lambda~.
\end{align}
 As is the case for all four-dimensional bosonic models with maximal depth, the action \eqref{2.7} is second order in derivatives and it takes the form
\begin{align}
S_{\text{Skeleton}}^{(2)}=\int \text{d}^4x \, e \, \mathfrak{C}^{\a(3)\ad}(h)\mathfrak{C}_{\a(3)\ad}(h)+\text{c.c.}~,\qquad \mathfrak{C}_{\a(3)\ad}(h)=\nabla_{(\a_1}{}^{\bd}h_{\a_2\a_3)\ad\bd}~.
\end{align}
This functional is not gauge invariant if the background Weyl tensor is non-vanishing, 
$C_{\a(4)} \neq 0$,  and one can show that its variation under \eqref{2.10} is equal to 
\begin{align}
\delta_{\lambda} S_{\text{Skeleton}}^{(2)}=2\int\text{d}^4x\, e \, \lambda\bigg\{C_{\a(3)}{}^{\d}\nabla_{\d\ad}\mathfrak{C}^{\a(3)\ad}(h)+2\mathfrak{C}^{\a(3)\ad}(h)\nabla_{\d\ad}C^{\d}{}_{\a(3)}\bigg\}+\text{c.c.}
\end{align}
However, there is one non-minimal primary term that can be added to $S_{\text{Skeleton}}^{(2)}$, 
\begin{align}
S_{\text{NM}}^{(2)}=\int\text{d}^4x &\, e \, h^{\a(2)\ad(2)}C_{\a(2)}{}^{\b(2)}h_{\b(2)\ad(2)} +{\rm c.c.}  \label{2.13}
\end{align}
 The variation of \eqref{2.13} under \eqref{2.10} is 
\begin{align} 
\delta_{\lambda}S_{\text{NM}}^{(2)}=\delta_{\lambda} S_{\text{Skeleton}}^{(2)}+\bigg( 2\int\text{d}^4x\, e \, \lambda B^{\a(2)\ad(2)}h_{\a(2)\ad(2)}+\text{c.c.}\bigg)~, \label{2.14}
\end{align}
where $B_{\a(2)\ad(2)}$ is the Bach tensor,
\begin{align}
 B_{\a(2) \ad(2)}=\nabla^{\b_1}{}_{(\ad_1}\nabla^{\b_2}{}_{\ad_2)}
  C_{\a(2) \b(2)}
  =\nabla_{(\a_1}{}^{\bd_1}  \nabla_{\a_2)}{}^{\bd_2}
 \bar{C}_{\ad(2) \bd(2) }=\bar{B}_{\a(2) \ad(2)}~.\label{2.15}
\end{align}
It follows that the primary action
\begin{align}
S_{\text{CHS}}^{(2)}&=S_{\text{Skeleton}}^{(2)}-S_{\text{NM}}^{(2)}\notag\\
&=\int \text{d}^4x \, e \, \bigg\{\mathfrak{C}^{\a(3)\ad}(h)\mathfrak{C}_{\a(3)\ad}(h)-  h^{\a(2)\ad(2)}C_{\a(2)}{}^{\b(2)}h_{\b(2)\ad(2)}\bigg\}+\text{c.c.} \label{2.16}
\end{align}
is gauge invariant in any Bach-flat background,
\begin{align}
\delta_{\lambda}S_{\text{CHS}}^{(2)}\bigg|_{B_{\a(2)\ad(2)}=0}=0~.
\end{align}
 
 To make contact with the existing literature, it is useful to present the degauged version of this model. The process of degauging consists of fixing the special conformal symmetry by gauging away the dilatation connection, $\mathfrak{b}_a=0$. After this, the special conformal connection may be shown to be proportional to the Schouten tensor, $\mathfrak{f}_{ab}=\frac{1}{2}P_{ab}$. The conformal covariant derivative then reduces to
 \begin{align}
 \mathfrak{b}_a=0~\qquad \implies \qquad \nabla_a=\mathcal{D}_a+\frac{1}{2}P_{a}{}^{b}K_{b}\label{degauge}
 \end{align} 
 where $\mathcal{D}_a=e_{a}{}^{m}\partial_m-\frac{1}{2}\omega_{a}{}^{bc}M_{bc}$ is the torsion-free Lorentz covariant derivative. 
 
 Upon degauging and converting to vector notation, the action \eqref{2.16} takes the form
 \begin{align}
 S_{\text{CHS}}^{(2)}=-8\int \text{d}^4x \, e \,\bigg\{&\mathcal{D}^ah^{bc}\mathcal{D}_{a}h_{bc}-\frac{4}{3}\mathcal{D}_{a}h^{ab}\mathcal{D}^{c}h_{bc}-2R_{ab}h^{ac}h_{c}{}^{b}+\frac{1}{6}Rh^{ab}h_{ab}\notag\\
 &+2C_{abcd}h^{ac}h^{bd}\bigg\}  \label{788.9}
 \end{align}
 where we have made use of \eqref{666} and the definition $h_{\a(2)\ad(2)}:=(\s^a)_{\a\ad}(\s^{b})_{\a\ad}h_{ab}$ for symmetric and traceless $h_{ab}$. It is invariant under the degauged transformations \eqref{2.10}, 
\begin{align}
\delta_{\lambda}h_{\a(2)\ad(2)}=\mathcal{D}_{(\a_1(\ad_1}\mathcal{D}_{\a_2)\ad_2)}\lambda-\frac{1}{2}R_{\a(2)\ad(2)}\lambda \label{45.87}
\end{align} 
 where $R_{\a\b\ad\bd}=(\sigma^a)_{\a\ad}(\sigma^b)_{\b\bd}\big(R_{ab}-\frac{1}{4}\eta_{ab}R\big)$ is the traceless part of the Ricci tensor. In vector notation the transformations \eqref{45.87} read
\begin{align}
\delta_{\lambda}h_{ab}=\big(\mathcal{D}_a\mathcal{D}_b-\frac{1}{2}R_{ab}\big)\lambda-\frac{1}{4}\eta_{ab}\big(\Box-\frac 12 R\big)\lambda~. \label{88.88}
\end{align} 
 
  The action \eqref{788.9} consists of two sectors that are independently invariant under Weyl transformations. Various combinations of these functionals were studied earlier in \cite{DeserN, DeserN2, LN, Sachs, EO } when trying to construct Weyl invariant second-order models for a symmetric traceless rank two tensor. However the question of gauge invariance was first raised in \cite{DeserN, DeserN2}, but only in the case of an (A)dS$_4$ background, where the last term in \eqref{788.9} is not present. Much later, the correct action \eqref{788.9} was proposed in \cite{BT2015}, but the authors considered only gauge transformations of the type
  \begin{align}
  \delta_{\lambda}h_{ab}=\big(\mathcal{D}_a\mathcal{D}_b-\frac{1}{4}\eta_{ab}\Box\big)\lambda~.\label{51.2}
  \end{align}
  Consequently, it was concluded that gauge invariance could only be upheld in Einstein spaces, where \eqref{88.88} and \eqref{51.2} coincide. 
It is important to emphasise that the equation of motion resulting from \eqref{788.9} was observed in \cite{DeserW6} to be invariant under the gauge transformations \eqref{88.88} in an arbitrary Bach-flat background. However the authors of \cite{DeserW6} were interested in coupling the model to conformal gravity, which lead to the conclusion that the system was inconsistent.


\section{Conformal spin-3 model in Bach-flat background}\label{section4}

The next case that we would like to analyse is the CMD
spin-3 field $h_{\a(3)\ad(3)}$, with $m=n=t=3$. Its gauge freedom is
\begin{align}
\delta_{\lambda}h_{\a(3)\ad(3)}=\nabla_{(\a_1(\ad_1}\nabla_{\a_2\ad_2}\nabla_{\a_3)\ad_3)}\lambda~.\label{2.18}
\end{align}
Both $h_{\a(3)\ad(3)}$ and $\lambda$ are primary and have Weyl weights
\begin{align}
\mathbb{D}h_{\a(3)\ad(3)}=h_{\a(3)\ad(3)}~,\qquad  \mathbb{D}\lambda=-2\lambda~.
\end{align}

 The conformal skeleton action 
\begin{align}
S_{\text{Skeleton}}^{(3)}=-\int\text{d}^4x \, e \,\mathfrak{C}^{\a(4)\ad(2)}(h)\mathfrak{C}_{\a(4)\ad(2)}(h)+ {\rm c.c.} ~, \qquad \mathfrak{C}_{\a(4)\ad(2)}(h)=\nabla_{(\a_1}{}^{\bd}h_{\a_2\a_3\a_4)\ad(2)\bd}\label{2.19}
\end{align}
has gauge variation equal to
\begin{align}
\delta_{\lambda}S_{\text{Skeleton}}^{(3)}=&\int\text{d}^4x \, e \, \lambda  \bigg\{8\mathfrak{C}^{\a(4)\ad(2)}(h)\nabla_{\a\ad}\nabla_{\g\ad}C_{\a(3)}{}^{\g}+4\nabla_{\ad}{}^{\g}C_{
\a(4)}\nabla_{\g\ad}\mathfrak{C}^{\a(4)\ad(2)} (h)\notag\\
&+16\nabla_{\a\ad}\mathfrak{C}^{\a(4)\ad(2)}(h)\nabla_{\g\ad}C_{\a(3)}{}^{\g}+\frac{16}{3}C_{\a(3)}{}^{\g}\nabla_{\a\ad}\nabla_{\g\ad}\mathfrak{C}^{\a(4)\ad(2)}(h)\bigg\}+\text{c.c.}
\end{align}
Once again, there is only one possible non-minimal primary term that is bilinear in $h_{\a(3)\ad(3)}$,
\begin{align}
S_{\text{NM}}^{(3)}=\int\text{d}^4x \, e \, h^{\g\a(2)\ad(3)}C_{\a(2)}{}^{\b(2)}h_{\b(2)\g\ad(3)}+\text{c.c.}~,\label{3.12}
\end{align}
and its gauge variation proves to be equal to
\begin{align}
\delta_{\lambda}S_{\text{NM}}^{(3)}=&-\frac{1}{2}\delta_{\lambda}S_{\text{Skeleton}}^{(3)} +\bigg(\int \text{d}^4x \, e \, \lambda\bigg\{-2\nabla^{\a\ad}B^{\a(2)\ad(2)}h_{\a(3)\ad(3)}-3B^{\a(2)\ad(2)}\nabla^{\a\ad}h_{\a(3)\ad(3)} \notag\\
&+\frac{8}{3}\bar{C}^{\ad(3)\bd}\nabla_{\b\bd}C^{\a(3)\b}h_{\a(3)\ad(3)}+\frac{8}{3}C^{\a(3)\b}\nabla_{\b\bd}\bar{C}^{\ad(3)\bd}h_{\a(3)\ad(3)}\notag\\
&+\frac{4}{3}C^{\a(3)\b}\bar{C}^{\ad(3)\bd}\nabla_{\b\bd}h_{\a(3)\ad(3)}\bigg\}+\text{c.c.}\bigg)~.\label{2.22}
\end{align}
We would like to point out that in deriving \eqref{2.22}, there is a nontrivial contribution arising from integration by parts, we discuss this technicality in more detail in the appendix. 

It follows that in a Bach-flat background, the deformed action
\begin{align}
S_{hh}^{(3)}=S_{\text{Skeleton}}^{(3)}+2S_{\text{NM}}^{(3)}~, \label{1234}
\end{align}
is gauge invariant only to first order in the background Weyl tensor, since 
\begin{align}
\delta_{\lambda}S_{hh}^{(3)}\bigg|_{B_{\a(2)\ad(2)}=0}= \frac{8}{3}\int \text{d}^4x \, e \, \lambda&\bigg\{2\bar{C}^{\ad(3)\bd}\nabla_{\b\bd}C^{\a(3)\b}h_{\a(3)\ad(3)}+2C^{\a(3)\b}\nabla_{\b\bd}\bar{C}^{\ad(3)\bd}h_{\a(3)\ad(3)}\notag\\
&+C^{\a(3)\b}\bar{C}^{\ad(3)\bd}\nabla_{\b\bd}h_{\a(3)\ad(3)}\bigg\}+\text{c.c.}
\label{2.24}
\end{align}
This is the best that one can achieve without making use of any extra fields.  

Our result \eqref{2.24} is analogous to the conclusion of Ref. \cite{GrigorievT, BeccariaT}
that the pure spin-3 action cannot be made gauge invariant beyond the first order in curvature. 
It was also conjectured in \cite{GrigorievT} (and later confirmed in \cite{BeccariaT}) that it might be possible to restore the spin-3
gauge invariance by introducing  a coupling   to a conformal spin-1 field. 
For the CMD spin-3 field, we are going to demonstrate 
that gauge invariance can indeed be restored by switching on a coupling to 
certain lower-spin fields.

To this aim, we introduce two lower-spin fields
$\chi_{\a(3)\ad}$ and $\vf_{\a(4)}$, along with their complex conjugates $\bar{\chi}_{\a\ad(3)}$ and $\bar{\vf}_{\ad(4)}$. They each carry the following conformal properties
\begin{subequations}\label{2.25}
\begin{align}
\mathbb{D}\chi_{\a(3)\ad}&=\chi_{\a(3)\ad}~,\qquad K_{\b\bd}\chi_{\a(3)\ad}=0~,\\
\mathbb{D}\vf_{\a(4)}&=0~,\qquad~~~~~~~~ K_{\b\bd}\vf_{\a(4)}=0~,
\end{align}
\end{subequations}
and are defined modulo gauge transformations 
\begin{subequations}\label{2.26}
\begin{align}
\delta_{\lambda}\chi_{\a(3)\ad}&=C_{\a(3)}{}^{\b}\nabla_{\b\ad}\lambda-2\nabla_{\b\ad}C_{\a(3)}{}^{\b}\lambda~,\\
\delta_{\lambda}\vf_{\a(4)}&=C_{\a(4)}\lambda~.
\end{align}
\end{subequations}
The right hand sides of \eqref{2.26} are fixed by the conformal properties \eqref{2.25}. To cancel the variation \eqref{2.22} we introduce the following couplings between the two lower-spin fields and $h$,
\begin{subequations}\label{2.27}
\begin{align}
S^{(3)}_{h\chi}&=\int\text{d}^4x\, e \, h^{\a(3)\ad(3)}\bar{C}_{\ad(3)}{}^{\bd}\chi_{\a(3)\bd}+\text{c.c.}~,\\
S^{(3)}_{h\bar{\vf}}&=\int\text{d}^4x\, e \, h^{\a(3)\ad(3)}\bigg\{C_{\a(3)}{}^{\b}\nabla_{\b\bd}\bar{\vf}_{\ad(3)}{}^{\bd}-3\nabla_{\b\bd}C_{\a(3)}{}^{\b}\bar{\vf}_{\ad(3)}{}^{\bd}\bigg\}+\text{c.c.}~,
\end{align} 
\end{subequations}
both of which are primary. Under \eqref{2.26} the functionals \eqref{2.27} transform as 
\begin{subequations}\label{2.28}
\begin{align}
\delta_{\lambda}S^{(3)}_{h\chi}=& \int\text{d}^4x \, e \,\bigg\{\bar{C}^{\ad(3)\bd}\chi^{\a(3)}{}_{\bd}\delta_{\lambda}h_{\a(3)\ad(3)} -\lambda\bigg[ 3\bar{C}^{\ad(3)\bd}\nabla_{\b\bd}C^{\a(3)\b}h_{\a(3)\ad(3)}\notag\\
&+C^{\a(3)\b}\nabla_{\b\bd}\bar{C}^{\ad(3)\bd}h_{\a(3)\ad(3)}+C^{\a(3)\b}\bar{C}^{\ad(3)\bd}\nabla_{\b\bd}h_{\a(3)\ad(3)}\bigg]\bigg\}+\text{c.c.} ~,\label{2.28a}\\
\delta_{\lambda}S^{(3)}_{h\bar{\vf}}=& \int\text{d}^4x \, e \,\bigg\{\delta_{\lambda}h^{\a(3)\ad(3)}\bigg[C_{\a(3)}{}^{\b}\nabla_{\b\bd}\bar{\vf}_{\ad(3)}{}^{\bd}-3\nabla_{\b\bd}C_{\a(3)}{}^{\b}\bar{\vf}_{\ad(3)}{}^{\bd}\bigg] \notag\\
&-\lambda\bigg[4\bar{C}^{\ad(3)\bd}\nabla_{\b\bd}C^{\a(3)\b}h_{\a(3)\ad(3)}+C^{\a(3)\b}\bar{C}^{\ad(3)\bd}\nabla_{\b\bd}h_{\a(3)\ad(3)}\bigg]\bigg\}+\text{c.c.}~ \label{2.28b}
\end{align}
\end{subequations}

Of course, the presence of the non-diagonal sector \eqref{2.27} forces us to introduce kinetic terms for each field so that we may cancel the first term in each of the variations \eqref{2.28a} and \eqref{2.28b}. It turns out that the appropriate kinetic actions take the form\footnote{The two terms on the right  of \eqref{2.29a} coincide modulo a total derivative. The same is true of  \eqref{2.29b}.}
\begin{subequations}\label{2.29}
\begin{align}
S^{(3)}_{\chi\bar{\chi}}=&~\frac{1}{2}\int\text{d}^4x\, e \, \chi^{\a(3)\ad}\nabla_{\a}{}^{\ad}\nabla_{\a}{}^{\ad}\bar{\chi}_{\a\ad(3)}+\text{c.c.}~,\label{2.29a}\\
S^{(3)}_{\vf\bar{\vf}}=&~\frac{1}{2}\int\text{d}^4x\, e \,\bar{\vf}^{\ad(4)}\nabla_{\ad}{}^{\a}\nabla_{\ad}{}^{\a}\nabla_{\ad}{}^{\a}\nabla_{\ad}{}^{\a}\vf_{\a(4)}+\text{c.c.}\label{2.29b}
\end{align}
\end{subequations}
They are both primary and prove to have the following gauge variations
\begin{subequations}\label{2.30}
\begin{align}
\delta_{\lambda}S^{(3)}_{\chi\bar{\chi}}=&-\int\text{d}^4x\, e \, \bigg\{ \bar{C}^{\ad(3)\bd}\chi^{\a(3)}{}_{\bd}\delta_{\lambda}h_{\a(3)\ad(3)}\notag\\
&+\lambda\bigg[\chi^{\a(3)\ad}\nabla_{\a}{}^{\ad}B_{\a(2)\ad(2)}+3B_{\a(2)\ad(2)}\nabla_{\a}{}^{\ad}\chi^{\a(3)\ad}\bigg]\bigg\}+\text{c.c.}~,\\
\delta_{\lambda}S^{(3)}_{\vf\bar{\vf}}=&\int\text{d}^4x\, e \,\bigg\{-\delta_{\lambda}h^{\a(3)\ad(3)}\bigg[C_{\a(3)}{}^{\b}\nabla_{\b\bd}\bar{\vf}_{\ad(3)}{}^{\bd}-3\nabla_{\b\bd}C_{\a(3)}{}^{\b}\bar{\vf}_{\ad(3)}{}^{\bd}\bigg]~~~~~~~~~~~~~~~~~~~~~~~~~~~~~~~~~~~~~~~~~~~~~~~~\notag\\
+\lambda\bigg[6B^{\a(2)\ad(2)}&\nabla_{\a}{}^{\ad}\nabla_{\a}{}^{\ad}\bar{\vf}_{\ad(4)}+8\nabla_{\a}{}^{\ad}B^{\a(2)\ad(2)}\nabla_{\a}{}^{\ad}\bar{\vf}_{\ad(4)}+3\bar{\vf}_{\ad(4)}\nabla_{\a}{}^{\ad}\nabla_{\a}{}^{\ad}B^{\a(2)\ad(2)}\bigg]\bigg\}+\text{c.c.}
\end{align}
\end{subequations}
From \eqref{2.28} and \eqref{2.30}, it follows that the conformal action
\begin{align}
S^{(3)}_{\text{CHS}}=&S_{hh}^{(3)}+\frac{16}{3}S^{(3)}_{\chi\bar{\chi}}-\frac{8}{3}S^{(3)}_{\vf\bar{\vf}}+\frac{16}{3}S^{(3)}_{h\chi}-\frac{8}{3}S^{(3)}_{h\bar{\vf}}~\label{2.31} \\
=&\int\text{d}^4x \, e \,\bigg\{-\mathfrak{C}^{\a(4)\ad(2)}(h)\mathfrak{C}_{\a(4)\ad(2)}(h)+2h^{\g\a(2)\ad(3)}C_{\a(2)}{}^{\b(2)}h_{\b(2)\g\ad(3)} \notag\\
&+\frac{8}{3}\chi^{\a(3)\ad}\nabla_{\a}{}^{\ad}\nabla_{\a}{}^{\ad}\bar{\chi}_{\a\ad(3)}-\frac{4}{3}\bar{\vf}^{\ad(4)}\nabla_{\ad}{}^{\a}\nabla_{\ad}{}^{\a}\nabla_{\ad}{}^{\a}\nabla_{\ad}{}^{\a}\vf_{\a(4)} +\frac{16}{3}h^{\a(3)\ad(3)}\bar{C}_{\ad(3)}{}^{\bd}\chi_{\a(3)\bd}\notag\\
&-\frac{8}{3}h^{\a(3)\ad(3)}\bigg[C_{\a(3)}{}^{\b}\nabla_{\b\bd}\bar{\vf}_{\ad(3)}{}^{\bd}-3\nabla_{\b\bd}C_{\a(3)}{}^{\b}\bar{\vf}_{\ad(3)}{}^{\bd}\bigg]\bigg\}+\text{c.c.}~, 
\end{align}
has gauge variation that is strictly proportional to the Bach tensor,
\begin{align}
\delta_{\lambda}S^{(3)}_{\text{CHS}}=&-\int\text{d}^4x\, e \, \lambda\bigg\{4\nabla^{\a\ad}B^{\a(2)\ad(2)}h_{\a(3)\ad(3)}+6B^{\a(2)\ad(2)}\nabla^{\a\ad}h_{\a(3)\ad(3)}+\frac{16}{3}\chi^{\a(3)\ad}\nabla_{\a}{}^{\ad}B_{\a(2)\ad(2)}\notag\\
&+16B_{\a(2)\ad(2)}\nabla_{\a}{}^{\ad}\chi^{\a(3)\ad}+16B^{\a(2)\ad(2)}\nabla_{\a}{}^{\ad}\nabla_{\a}{}^{\ad}\bar{\vf}_{\ad(4)}+\frac{64}{3}\nabla_{\a}{}^{\ad}B^{\a(2)\ad(2)}\nabla_{\a}{}^{\ad}\bar{\vf}_{\ad(4)}\notag\\
&+8\bar{\vf}_{\ad(4)}\nabla_{\a}{}^{\ad}\nabla_{\a}{}^{\ad}B^{\a(2)\ad(2)}\bigg\}+\text{c.c.}
\end{align}
It is therefore gauge invariant when restricted to a Bach-flat background,
\begin{align}
\delta_{\lambda}S_{\text{CHS}}^{(3)}\bigg|_{B_{\a(2)\ad(2)}=0}=0~.
\end{align}
Due to the presence of the kinetic terms, the action \eqref{2.31} does not reduce to \eqref{2.19} in the conformally flat limit, but rather to 
\begin{align}
S_{\text{CHS}}^{(3)}\bigg|_{C_{abcd}=0}=S_{\text{Skeleton}}^{(3)}+\frac{16}{3}S^{(3)}_{\chi\bar{\chi}}-\frac{8}{3}S^{(3)}_{\vf\bar{\vf}}~.
\end{align}

Finally, it is of interest to provide the degauged version of the pure spin-3 sector \eqref{1234} in vector notation. It may be shown to be
\begin{align}
 S_{hh}^{(3)}=8\int \text{d}^4x \, e \,\bigg\{&\mathcal{D}^ah_{acd}\mathcal{D}^{b}h_{b}{}^{cd}-2\mathcal{D}_{a}h_{bcd}\mathcal{D}^{a}h^{bcd}+6R_{ab}h^{acd}h_{cd}{}^{b}-\frac{2}{3}Rh^{abc}h_{abc}\notag\\
 &-8C_{abcd}h^{acf}h_{f}{}^{bd}\bigg\}  \label{788}
 \end{align}
where we have made use of the definition $h_{\a(3)\ad(3)}:=(\s^a)_{\a\ad}(\s^b)_{\a\ad}(\s^c)_{\a\ad}h_{abc}$ for symmetric and traceless $h_{abc}$. The gauge transformations \eqref{2.18} are then equivalent to
\begin{align}
\delta_{\lambda}h_{abc}=\bigg(\mathcal{D}_{(a}\mathcal{D}_{b}\mathcal{D}_{c)}-2R_{(ab}\mathcal{D}_{c)}-\mathcal{D}_{(a}R_{bc)}\bigg)\lambda+\eta_{(ab}\bigg(R_{c)}{}^{d}\mathcal{D}_{d}+\frac{1}{3}R\mathcal{D}_{c)}+\frac{1}{3}\mathcal{D}_{c)}R-\frac{1}{2}\mathcal{D}_{c)}\Box\bigg)\lambda~.
\end{align}
The conversion of the lower-spin sectors in \eqref{2.31} is a straightforward but tedious matter and will be omitted as the final expressions are not illuminating.


\section{Conformal spin-5/2 model in Bach-flat background}\label{section5}

Maximal depth fermionic models (half-integer spin) differ from  their bosonic counterparts in that their skeletons \eqref{2.7} are all third order in derivatives. This makes 
extending them to Bach-flat backgrounds technically more challenging, but conceptually there is no difference. In particular, as we show below, lower-spin fields must also be introduced to render the spin-$5/2$ system gauge invariant beyond first order in the Weyl curvature.

The CMD spin-5/2 field $\psi_{\a(3)\ad(2)}$
(which corresponds to  $m-1=n=t=2$ in the notation of section \ref{section2})
  is defined modulo depth two gauge transformations
\begin{align}
\delta_{\lambda}\psi_{\a(3)\ad(2)}=\nabla_{(\a_1(\ad_1}\nabla_{\a_2\ad_2)}\lambda_{\a_3)}~.\label{3.29}
\end{align}
Both $\psi_{\a(3)\ad(2)}$ and $\lambda_{\a}$ are primary and carry Weyl weights
\begin{align}
\mathbb{D}\psi_{\a(3)\ad(2)}=\frac{1}{2}\psi_{\a(3)\ad(2)}~,\qquad \mathbb{D}\lambda_{\a}=-\frac{3}{2}\lambda_{\a}~.
\end{align}

The skeleton sector \eqref{2.7},
\begin{align}
S_{\text{Skeleton}}^{(5/2)}[\psi,\bar{\psi}]={\rm i}\int \text{d}^4x \, e \,\hat{\mathfrak{C}}^{\a(4)\ad}(\psi)\check{\mathfrak{C}}_{\a(4)\ad}(\bar\psi) +\text{c.c.}~,\label{88.00}
\end{align}
is composed of the two generalised Weyl tensors
\begin{align}
\hat{\mathfrak{C}}_{\a(4)\ad}(\psi)=\nabla_{(\a_1}{}^{\bd}\psi_{\a_2\a_3\a_4)\ad\bd}~,\qquad \check{\mathfrak{C}}_{\a(4)\ad}(\bar\psi)=\nabla_{(\a_1}{}^{\bd}\nabla_{\a_2}{}^{\bd}\bar\psi_{\a_3\a_4)\ad\bd(2)}~.
\end{align}
Under the transformation \eqref{3.29} it varies as
\begin{align}
\delta_{\lambda}S_{\text{Skeleton}}^{(5/2)}&={\rm i}\int\text{d}^4x\, e \, \bigg\{\lambda^{\a}\bigg[\frac{5}{2}C^{\g\b(3)}\nabla_{\g}{}^{\bd}\check{\mathfrak{C}}_{\a\b(3)\bd}(\bar\psi)+3\nabla_{\g}{}^{\bd}C^{\g\b(3)}\check{\mathfrak{C}}_{\a\b(3)\bd}(\bar\psi)-\frac{3}{2}C^{\b(4)}\nabla_{\a}{}^{\bd}\check{\mathfrak{C}}_{\b(4)\bd}(\bar\psi)\notag\\
&-\nabla_{\a}{}^{\bd}C^{\b(4)}\check{\mathfrak{C}}_{\b(4)\bd}(\bar\psi)\bigg] - \frac{1}{3}\bar\lambda_{\ad}\bigg[-\nabla^{\d\dd}C^{\b(4)}\nabla_{\d\dd}\hat{\mathfrak{C}}_{\b(4)}{}^{\ad}(\psi)+\Box C^{\b(4)}\hat{\mathfrak{C}}_{\b(4)}{}^{\ad}(\psi)\notag\\
&+6\nabla_{\g}{}^{\dd}C^{\g\b(3)}\nabla_{\dd}{}^{\b}\hat{\mathfrak{C}}_{\b(4)}{}^{\ad}(\psi)+2\nabla^{\b\bd}C^{\b(3)\g}\nabla_{\g}{}^{\ad}\hat{\mathfrak{C}}_{\b(4)\bd}(\psi)+5\nabla^{\b\bd}\nabla_{\g}{}^{\ad}C^{\g\b(3)}\hat{\mathfrak{C}}_{\b(4)\bd}(\psi)\notag\\
&+10\nabla_{\g}{}^{\ad}C^{\g\b(3)}\nabla^{\b\bd}\hat{\mathfrak{C}}_{\b(4)\bd}(\psi)+4 C^{\b(4)}\Box\hat{\mathfrak{C}}_{\b(4)}{}^{\ad}(\psi)+4C^{\b(3)\g}\nabla_{\g}{}^{\ad}\nabla^{\b\bd}\hat{\mathfrak{C}}_{\b(4)\bd}(\psi)\notag\\
&-15C^{\b(2)\d(2)}C_{\d(2)}{}^{\b(2)}\hat{\mathfrak{C}}_{\b(4)}{}^{\ad}(\psi)\bigg] \bigg\} +\text{c.c.}
\end{align}
Unlike the previous bosonic models, for spin-$5/2$ there is a family of non-minimal primary counter-terms, which is generated by the following two functionals\footnote{There are also two more functionals of the form ${\rm i}\int \text{d}^4x \, e \, \psi^{\a(3)\ad(2)}\mathfrak{J}_{\a(3)\ad(2)}(\bar\psi)+\text{c.c.}$, where $\mathfrak{J}_{\a(3)\ad(2)}(\bar\psi)$ is a composite primary field depending on $\bar{C}_{\ad(4)}$ and $\bar\psi_{\a(2)\ad(3)}$. However they prove to be equivalent to \eqref{3.34} modulo total derivatives. }
\begin{subequations}\label{3.34}
\begin{align}
S_{\text{NM}}^{(5/2)}&={\rm i}\int\text{d}^4x\, e \, \psi^{\a(3)\ad(2)}\bigg\{-\frac{5}{4}C_{\a(3)}{}^{\b}\nabla^{\b\bd}\bar{\psi}_{\b(2)\bd\ad(2)}+\nabla^{\b\bd}C_{\a(3)}{}^{\b}\bar{\psi}_{\b(2)\bd\ad(2)}\notag \\
&\phantom{={\rm i}\int\text{d}^4x\, e \, \psi^{\a(3)\ad(2)}\bigg\{ }~+3C_{\a(2)}{}^{\b(2)}\nabla_{\a}{}^{\bd}\bar\psi_{\b(2)\bd\ad(2)} \bigg\} +\text{c.c.}~, \label{3.34a}\\
\widetilde{S}_{\text{NM}}^{(5/2)}&={\rm i}\int\text{d}^4x\, e \, \psi^{\a(3)\ad(2)}\bigg\{C_{\a(3)}{}^{\b}\nabla^{\b\bd}\bar{\psi}_{\b(2)\bd\ad(2)}-2\nabla^{\b\bd}C_{\a(3)}{}^{\b}\bar{\psi}_{\b(2)\bd\ad(2)}\notag\\
&\phantom{={\rm i}\int\text{d}^4x\, e \, \psi^{\a(3)\ad(2)}\bigg\{ }~+3\nabla_{\a}{}^{\bd}C_{\a(2)}{}^{\b(2)}\bar\psi_{\b(2)\bd\ad(2)} \bigg\} +\text{c.c.}
\end{align}
\end{subequations}
The overall coefficients in \eqref{3.34} are chosen so that their variations may cancel that of \eqref{88.00}.

To first order in the Weyl tensor, it may be shown that any linear combination of the two functionals, $a_1S_{\text{NM}}^{(5/2)}+a_2\widetilde{S}_{\text{NM}}^{(5/2)}$, with $a_2\neq 0$ will have gauge variation not proportional to that of $S_{\text{Skeleton}}^{(5/2)}$. Therefore it suffices to consider only the first structure \eqref{3.34a}. Indeed, its gauge variation may be shown to be
\begin{align}
\delta_{\lambda}S_{\text{NM}}^{(5/2)}&=\delta_{\lambda}S_{\text{Skeleton}}^{(5/2)}+\bigg(\text{i}\int\text{d}^4x\, e \, \bigg\{ \lambda^{\a}\bigg[\frac{3}{4}B_{\a}{}^{\b\bd(2)}\nabla^{\b\bd}\bar\psi_{\b(2)\bd(3)}+\nabla^{\b\bd}B_{\a}{}^{\b\bd(2)}\bar\psi_{\b(2)\bd(3)}\notag\\
&+B^{\b(2)\bd(2)}\nabla_{\a}{}^{\bd}\bar\psi_{\b(2)\bd(3)}-\frac{7}{4}\bar C^{\gd\bd(3)}\nabla_{\g\gd}C_{\a}{}^{\g\b(2)}\bar\psi_{\b(2)\bd(3)}-\frac{13}{8}C_{\a}{}^{\g\b(2)}\nabla_{\g\gd}\bar C^{\gd\bd(3)}\bar\psi_{\b(2)\bd(3)}\notag\\
&-\frac{9}{8}\bar C^{\gd\bd(3)}C_{\a}{}^{\g\b(2)}\nabla_{\g\gd}\bar\psi_{\b(2)\bd(3)}\bigg]-\bar\lambda_{\ad}\bigg[\nabla^{\b\bd}B^{\b(2)\bd\ad}\psi_{\b(3)\bd(2)}+\frac{3}{4}B^{\b(2)\bd\ad}\nabla^{\b\bd}\psi_{\b(3)\bd(2)} \notag\\
&+B^{\b(2)\bd(2)}\nabla^{\b\ad}\psi_{\b(3)\bd(2)}-\frac{49}{24}\bar C^{\ad\gd\bd(2)}\nabla_{\g\gd}C^{\g\b(3)}\psi_{\b(3)\bd(2)}-\frac{4}{3} C^{\g\b(3)}\nabla_{\g\gd}\bar C^{\ad\gd\bd(2)}\psi_{\b(3)\bd(2)}\notag\\
&-\frac{9}{8}\bar C^{\ad\gd\bd(2)}C^{\g\b(3)}\nabla_{\g\gd}\psi_{\b(3)\bd(2)}\bigg]\bigg\}+\text{c.c.}\bigg)~.
\end{align}

We see that once again, using just the spin-$5/2$ field, gauge invariance can only be controlled to first order in the Weyl tensor. To go beyond this order we need to introduce two lower-spin fields\footnote{In principle one could also consider the field $\rho_{\a(4)\ad}$, which has the same conformal properties as $\vf_{\a(3)}$ but has the gauge transformation $\delta_{\lambda}\rho_{\a(4)\ad}=C_{\a(4)}\bar\lambda_{\ad}$. However this field turns out to be unnecessary in the construction.} 
$\chi_{\a(2)\ad}$
and $\vf_{\a(3)}$.
They possess the conformal properties
\begin{subequations}
\begin{align}
\mathbb{D}\chi_{\a(2)\ad}&=\frac{3}{2}\chi_{\a(2)\ad}~,\qquad K_{\b\bd}\chi_{\a(2)\ad}=0~,\\
\mathbb{D}\vf_{\a(3)}&=\frac{1}{2}\vf_{\a(3)}~,\qquad~~~ K_{\b\bd}\vf_{\a(3)}=0~,
\end{align}
\end{subequations}
and are defined modulo the gauge transformations 
\begin{subequations}
\begin{align}
\delta_{\lambda}\chi_{\a(2)\ad}&=C_{\a(2)}{}^{\b(2)}\nabla_{\b\ad}\lambda_{\b}-\nabla_{\b\ad}C_{\a(2)}{}^{\b(2)}\lambda_{\b}~,\\
\delta_{\lambda}\vf_{\a(3)}&=C_{\a(3)}{}^{\b}\lambda_{\b}~.
\end{align}
\end{subequations}

The primary couplings between these fields and the spin-$5/2$ field take the form
\begin{subequations}\label{6.90}
\begin{align}
S^{(5/2)}_{\bar\psi\chi}&= \text{i}\int\text{d}^4x\, e \,\bar\psi^{\a(2)\ad(3)}\bar C_{\ad(3)}{}^{\bd}\chi_{\a(2)\bd}+\text{c.c.}~, \\
S^{(5/2)}_{\psi\bar\vf}&= \text{i}\int\text{d}^4x\, e \,\psi^{\a(3)\ad(2)}\bigg\{C_{\a(3)}{}^{\g}\nabla_{\g}{}^{\bd}\bar\vf_{\ad(2)\bd}-2\nabla_{\g}{}^{\bd}C_{\a(3)}{}^{\g}\bar\vf_{\ad(2)\bd}\bigg\}+\text{c.c.}
\end{align}
\end{subequations}
Their gauge variations may be shown to be 
\begin{subequations}\label{3.39}
\begin{align}
\delta_{\lambda}S^{(5/2)}_{\bar\psi\chi}&= \text{i}\int\text{d}^4x\, e \, \bigg\{ \lambda^{\a}\bigg[C_{\a}{}^{\g\b(2)}\bar C^{\gd\bd(3)}\nabla_{\g\gd}\bar\psi_{\b(2)\bd(3)}+C_{\a}{}^{\g\b(2)}\nabla_{\g\gd}\bar C^{\gd\bd(3)}\bar\psi_{\b(2)\bd(3)}\notag\\ 
&\phantom{= \text{i}\int\text{d}^4x\, e \, \bigg\{}+2\bar C^{\gd\bd(3)}\nabla_{\g\gd}C_{\a}{}^{\g\b(2)}\bar\psi_{\b(2)\bd(3)}\bigg]-\bar\lambda_{\ad}\bigg[\bar{C}^{\ad\bd\gd(2)}\nabla_{\gd}{}^{\b}\nabla_{\gd}{}^{\b}\chi_{\b(2)\bd} \notag\\
&\phantom{= \text{i}\int\text{d}^4x\, e \, \bigg\{}+2\nabla_{\gd}{}^{\b}\bar{C}^{\ad\bd\gd(2)}\nabla_{\gd}{}^{\b}\chi_{\b(2)\bd} +B^{\b(2)\ad\bd}\chi_{\b(2)\bd}\bigg]\bigg\}+\text{c.c.}~,\\
\delta_{\lambda}S^{(5/2)}_{\psi\bar\vf}&= \text{i}\int\text{d}^4x\, e \, \bigg\{\bar\lambda_{\ad}\bigg[C^{\g\b(3)}\bar C^{\ad\gd\bd(2)}\nabla_{\g\gd}\psi_{\b(3)\bd(2)}+3\bar C^{\ad\gd\bd(2)}\nabla_{\g\gd}C^{\g\b(3)}\psi_{\b(3)\bd(2)}\bigg]\notag\\
&\phantom{= \text{i}\int\text{d}^4x\, e \, \bigg\{}+\lambda^{\a}\bigg[C_{\a}{}^{\b(3)}\nabla_{\b}{}^{\bd}\nabla_{\b}{}^{\bd}\nabla_{\b}{}^{\bd}\bar\vf_{\bd(3)}-3B_{\a}{}^{\g\bd(2)}\nabla_{\g}{}^{\bd}\bar\vf_{\bd(3)}\notag\\
&\phantom{= \text{i}\int\text{d}^4x\, e \, \bigg\{}-2\nabla_{\g}{}^{\bd}B_{\a}{}^{\g\bd(2)}\bar\vf_{\bd(3)}\bigg]\bigg\} +\text{c.c.}
\end{align}
\end{subequations}
The kinetic actions required to cancel the variations in \eqref{3.39} proportional to the lower-spin fields are
\begin{subequations}
\begin{align}
S^{(5/2)}_{\chi\bar\chi}&= \frac{\text{i}}{2}\int\text{d}^4x\, e \,\chi^{\a(2)\ad}\nabla_{\a}{}^{\ad}\bar\chi_{\a\ad(2)}+\text{c.c.}~, \\
S^{(5/2)}_{\vf\bar{\vf}}&=\frac{\text{i}}{2}\int\text{d}^4x\, e \,\bar{\vf}^{\ad(3)}\nabla_{\ad}{}^{\a}\nabla_{\ad}{}^{\a}\nabla_{\ad}{}^{\a}\vf_{\a(3)}+\text{c.c.}
\end{align}
\end{subequations} 
They are both primary and prove to have the gauge transformations
\begin{subequations}
\begin{align}
\delta_{\lambda}S^{(5/2)}_{\chi\bar\chi}&=-\text{i}\int\text{d}^4x\, e \,\bar\lambda_{\ad}\bigg\{2\nabla_{\gd}{}^{\b}\bar C^{\ad\bd\gd(2)}\nabla_{\gd}{}^{\b}\chi_{\b(2)\bd}+\bar C^{\ad\bd\gd(2)}\nabla_{\gd}{}^{\b}\nabla_{\gd}{}^{\b}\chi_{\b(2)\bd}\bigg\}+\text{c.c.}~,\\
\delta_{\lambda}S^{(5/2)}_{\vf\bar{\vf}}&=-\text{i}\int\text{d}^4x\, e \,\lambda^{\a}C_{\a}{}^{\b(3)}\nabla_{\b}{}^{\bd}\nabla_{\b}{}^{\bd}\nabla_{\b}{}^{\bd}\bar\vf_{\bd(3)}+\text{c.c.}
\end{align}
\end{subequations} 
It follows that the action
\begin{align}
S_{\text{CHS}}^{(5/2)}&=S^{(5/2)}_{\text{Skeleton}}-S^{(5/2)}_{\text{NM}}-\frac{37}{24}S^{(5/2)}_{\bar\psi\chi}+\frac{17}{24}S^{(5/2)}_{\psi\bar\vf}+\frac{37}{24}S^{(5/2)}_{\chi\bar\chi}+\frac{17}{24}S^{(5/2)}_{\vf\bar\vf} \label{3.43}\\[2ex]
&=\text{i}\int\text{d}^4x\, e \,\bigg\{\hat{\mathfrak{C}}^{\a(4)\ad}(\psi)\check{\mathfrak{C}}_{\a(4)\ad}(\bar\psi)+ \psi^{\a(3)\ad(2)}\bigg[\frac{5}{4}C_{\a(3)}{}^{\b}\nabla^{\b\bd}\bar{\psi}_{\b(2)\bd\ad(2)}\notag\\
&-\nabla^{\b\bd}C_{\a(3)}{}^{\b}\bar{\psi}_{\b(2)\bd\ad(2)}-3C_{\a(2)}{}^{\b(2)}\nabla_{\a}{}^{\bd}\bar\psi_{\b(2)\bd\ad(2)}\bigg]+\psi^{\a(3)\ad(2)}\bigg[\frac{17}{24}C_{\a(3)}{}^{\g}\nabla_{\g}{}^{\bd}\bar\vf_{\ad(2)\bd}\notag\\
&-\frac{17}{12}\nabla_{\g}{}^{\bd}C_{\a(3)}{}^{\g}\bar\vf_{\ad(2)\bd} \bigg]-\frac{37}{24}\bar\psi^{\a(2)\ad(3)}\bar C_{\ad(3)}{}^{\bd}\chi_{\a(2)\bd}+\frac{37}{48}\chi^{\a(2)\ad}\nabla_{\a}{}^{\ad}\bar\chi_{\a\ad(2)}\notag\\
&+\frac{17}{48}\bar{\vf}^{\ad(3)}\nabla_{\ad}{}^{\a}\nabla_{\ad}{}^{\a}\nabla_{\ad}{}^{\a}\vf_{\a(3)}\bigg\}+\text{c.c.}~,
\end{align}
has gauge variation that is strictly proportional to the Bach tensor
\begin{align}
\delta_{\lambda}S^{(5/2)}_{\text{CHS}}&=-\text{i}\int\text{d}^4x\, e \,\bigg\{ \lambda^{\a}\bigg[\frac{3}{4}B_{\a}{}^{\b\bd(2)}\nabla^{\b\bd}\bar\psi_{\b(2)\bd(3)}+\nabla^{\b\bd}B_{\a}{}^{\b\bd(2)}\bar\psi_{\b(2)\bd(3)}\notag\\
&+B^{\b(2)\bd(2)}\nabla_{\a}{}^{\bd}\bar\psi_{\b(2)\bd(3)}+\frac{17}{8}B_{\a}{}^{\g\bd(2)}\nabla_{\g}{}^{\bd}\bar\vf_{\bd(3)}+\frac{17}{12}\nabla_{\g}{}^{\bd}B_{\a}{}^{\g\bd(2)}\bar\vf_{\bd(3)}\bigg]\notag\\
&-\bar\lambda_{\ad}\bigg[\nabla^{\b\bd}B^{\b(2)\bd\ad}\psi_{\b(3)\bd(2)}+\frac{3}{4}B^{\b(2)\bd\ad}\nabla^{\b\bd}\psi_{\b(3)\bd(2)} \notag\\
&+B^{\b(2)\bd(2)}\nabla^{\b\ad}\psi_{\b(3)\bd(2)}+\frac{37}{24}B^{\b(2)\bd\ad}\chi_{\b(2)\bd}\bigg]\bigg\}+\text{c.c.}
\end{align}
It is therefore gauge invariant in any Bach-flat background
\begin{align}
\delta_{\lambda}S^{(5/2)}_{\text{CHS}}\bigg|_{B_{\a(2)\ad(2)}=0}=0~,
\end{align}
and has the conformally flat limit
\begin{align}
S^{(5/2)}_{\text{CHS}}\bigg|_{C_{abcd}=0}=S^{(5/2)}_{\text{Skeleton}}+\frac{37}{24}S^{(5/2)}_{\chi\bar\chi}+\frac{17}{24}S^{(5/2)}_{\vf\bar\vf}~.
\end{align}


\section{Concluding comments}\label{section6}

In this paper we have provided the first two consistent models for conformal higher-spin fields propagating on four-dimensional Bach-flat backgrounds. To mitigate technical difficulties, we have considered the simpler problem associated with conformal fields of maximal depth.  We have found that when $s>2$, certain lower-spin fields are required in order to restore gauge invariance beyond first order in the background curvature. We have explored only the $s=5/2$ and $s=3$ cases in detail, but expect that similar models can also be constructed for higher-spins. In such models it is likely that the number of lower-spin fields required will increase. We plan to revisit these issues in the future. 

It is important to point out that the lower-spin fields are not gauge fields in the sense that they cannot have their own standard gauge transformations of the type \eqref{gt}. This is because their conformal weights differ to that prescribed by \eqref{weight}. This property is an artefact of the higher-depth nature of the CHS fields under consideration, and is one of the main differences to the model analysed in \cite{BeccariaT}. In Ref. \cite{BeccariaT}, the authors considered a coupling between the spin $s=1$ and $s=3$ conformal gauge fields with depth $t=1$. In that context, it is possible to consistently entangle their gauge symmetries whilst simultaneously preserving the conformal symmetry.

 To illustrate this, let us denote these fields by $h^{(1)}_{\a\ad}$ and $h^{(1)}_{\a(3)\ad(3)}$. According to \eqref{weight}, if they are to be conformal and defined modulo depth 1 gauge transformations \eqref{gt}, then their conformal weights are fixed to be 1 and $-1$ respectively.
 These restrictions allow for entangled gauge transformations such 
 as\footnote{ By inspection of the weights, it is not possible for the gauge transformation of the spin-3 field to involve the spin-1 gauge parameter.} 
\begin{subequations}\label{6.1}
\begin{align}
\delta_{\lambda}h^{(1)}_{\a(3)\ad(3)}&=\nabla_{(\a_1(\ad_1}\lambda^{(1)}_{\a_2\a_3)\ad_2\ad_3)}~,\\
\delta_{\lambda}h^{(1)}_{\a\ad} &=\nabla_{\a\ad}\lambda^{(1)}+
\Big[ C_{\a}{}^{\b(2)\g}\nabla_{\g}{}^{\bd}\lambda^{(1)}_{\b(2)\bd\ad}-3\nabla_{\g}{}^{\bd}C_{\a}{}^{\b(2)\g}\lambda^{(1)}_{\b(2)\bd\ad} +{\rm c.c.} \Big]~.
\end{align}
\end{subequations}
As per usual, the right hand sides are determined by the primary condition.

In contrast, for the maximal depth spin-3 field $h^{(3)}_{\a(3)\ad(3)}$ to be conformal and defined modulo the gauge transformations \eqref{2.18}, its conformal weight must be equal to  1 (and so its gauge parameter $\lambda^{(3)}$ has weight $-2$). There is no possible way to deform the gauge transformations of the spin-1 field (or spin-2 for that matter) to include the parameter $\lambda^{(3)}$ in a way that preserves its conformal symmetry and index structure. Thus we are forced to introduce exotic fields such as $\chi_{\a(3)\ad}$ and $\vf_{\a(4)}$, Eq. \eqref{2.26}, which are not technically gauge fields and whose physical meaning is obscure. 

So far, we have confined our attention to gauge fields with spin less than or equal to three. A natural question to ask is 
whether
similar lower-spin couplings 
are expected
to be necessary in the construction of gauge-invariant models for fields $\phi^{(t)}_{\a(m)\ad(n)}$ with $m+n>6$. For minimal depth gauge fields ($t=1$), it is 
not hard
to work out a higher-spin generalisation of 
the gauge transformation \eqref{6.1}.
  In the bosonic case, with $m=n=s>3$, it takes the form
\begin{subequations}\label{6.2}
\begin{align}
\delta_{\lambda}h^{(1)}_{\a(s)\ad(s)}&=\nabla_{(\a_1(\ad_1}\lambda^{(1)}_{\a_2\dots\a_s)\ad_2\dots\ad_s)}~,\\
\delta_{\lambda}h^{(1)}_{\a(s-2)\ad(s-2)}&=\nabla_{(\a_1(\ad_1}\lambda^{(1)}_{\a_2\dots\a_{s-2})\ad_2\dots\ad_{s-2})}+\bigg[\big(sa_1-2a_2\big)C_{(\a_1}{}^{\b(2)\g}\nabla_{|\g|}{}^{\bd}\lambda^{(1)}_{\a_2\dots\a_{s-2})\b(2)\bd\ad(s-2)}\notag\\
&\phantom{=}-\big(s^2a_1-3(s-1)a_2\big)\nabla_{\g}{}^{\bd}C_{(\a_1}{}^{\b(2)\g}\lambda^{(1)}_{\a_2\dots\a_{s-2})\b(2)\bd\ad(s-2)}\notag\\
&\phantom{=}+\hf (s-3)a_1C_{(\a_1\a_2}{}^{\b(2)}\nabla^{\b\bd}\lambda^{(1)}_{\a_3\dots\a_{s-2})\b(3)\bd\ad(s-2)}\notag\\
&\phantom{=}+\hf(s-3)a_2\nabla^{\b\bd}C_{(\a_1\a_2}{}^{\b(2)}\lambda^{(1)}_{\a_3\dots\a_{s-2})\b(3)\bd\ad(s-2)}+\text{c.c.}\bigg]~, \label{6.2b}
\end{align}
\end{subequations}
and this may be seen to be equivalent to \eqref{6.1} in the $s=3$ case. 
The presence of the two free parameters $a_1,a_2\in\mathbb{R}$ in \eqref{6.2b} 
signals that the greater the spin, the more freedom there is to entangle the gauge transformations. 
Actually, analogues of \eqref{6.2b} also exist for lower-spin gauge fields $h^{(1)}_{\a(s-s')\ad(s-s')}$ with $1\leq s' \leq s-1$. It is even conceivable to couple the parent field to a non-gauge field in a similar fashion to the maximal depth models presented 
in this paper.\footnote{Recently we have constructed a gauge-invariant model for the
 conformal gauge field  $\f^{(1)}_{\a(3)\ad}$ in an arbitrary Bach-flat background \cite{KPR}, 
 and the action involves a non-gauge field, similar to the  CMD models studied above.}
 Therefore, it seems reasonable to expect that the number of lower-spin couplings required will increase with the spin of the parent field.
However, explicit calculations are needed in order to understand which 
 lower-spin field(s) will be necessary. So far, no calculations in Bach-flat backgrounds 
 have been carried out  for minimal depth gauge fields $h^{(1)}_{\a(s)\ad(s)}$ 
 with $s>3$.


In conclusion we point out that the lower-spin fields in the CMD $s=5/2$ and $s=3$ models 
contribute to the Weyl anomalies. It would be interesting to compute these contributions.


 \noindent
{\bf Acknowledgements:}\\
Conversations with Misha Vasiliev are gratefully acknowledged.
A preliminary version of this work was reported by one of us (MP) at the APCTP Workshop 
``Higher Spin Gravity: Chaotic, Conformal and Algebraic Aspects''
(Pohang, South Korea). 
We are grateful to the organisers of the workshop  
for the fantastic scientific atmosphere and generous support.
The work of SMK is supported in part by the Australian 
Research Council, project No. DP160103633.
The work of MP is supported by the Hackett Postgraduate Scholarship UWA,
under the Australian Government Research Training Program. 


\appendix

\section{Aspects of integration by parts}

In general, integrating by parts in conformal space is nontrivial because the conformal covariant derivative carries extra connections that give non-vanishing contributions from total derivatives.\footnote{We refer the reader to appendix D of Ref. \cite{KP2} for a more detailed discussion on this technical issue.} However, under special conditions, which in practice are usually met, we may follow the usual procedure and ignore any total derivatives that arise. These conditions are not met for the non-minimal action \eqref{3.12} and so below we elaborate on how integration by parts works in this case. 

In what follows we drop the `$+$ c.c.' for simplicity. The gauge variation of \eqref{3.12} is 
\begin{align}
\delta_{\lambda}S^{(3)}_{\text{NM}}&=2\int\text{d}^4x \, e \, \nabla_{(\a_1\ad_1}\nabla_{\a_2\ad_2}\nabla_{\g)\ad_3}\lambda C^{\a(2)\b(2)}h_{\b(2)}{}^{\g\ad(3)} \notag\\
&=\frac{1}{3}\int\text{d}^4x \, e \, 
\bigg[6\nabla_{\g\ad_1}\nabla_{\a_1\ad_2}\nabla_{\a_2\ad_3}\lambda-4\ve_{\a_1\g}\bar{C}_{\ad(3)}{}^{\gd}\nabla_{\a_2\gd}\lambda-8\ve_{\a_1\g}\nabla_{\a_2\gd}\bar{C}_{\ad(3)}{}^{\gd}\lambda\bigg]\notag\\
&
\qquad \qquad \qquad 
\times C^{\a(2)\b(2)}h_{\b(2)}{}^{\g\ad(3)}
\notag\\
&= \mathcal{I}_{\text{Total}}+\frac{1}{3}\int\text{d}^4x \, e \, \lambda\bigg\{-6\nabla_{\a_1\ad_1}\nabla_{\a_2\ad_2}\nabla_{\g\ad_3}\big[C^{\a(2)\b(2)}h_{\b(2)}{}^{\g\ad(3)}\big]\notag\\
&~~~~~~~~~~~~~~~~+4\nabla_{\g\gd}\big[\bar{C}_{\ad(3)}{}^{\gd}C^{\g\b(3)}h_{\b(3)}{}^{\ad(3)}\big]-8\nabla_{\g\gd}\bar{C}_{\ad(3)}{}^{\gd}C^{\g\b(3)}h_{\b(3)}{}^{\ad(3)}\bigg\}~. 
\end{align}
Here $\mathcal{I}_{\text{Total}}$ represents the total derivative that arises in moving from the second to third line,
\begin{align}
\mathcal{I}_{\text{Total}}=\frac{1}{3}\int\text{d}^4x \, e \,\nabla^{\a\ad}\mathcal{Z}_{\a\ad}    ~,\label{A.2}
\end{align}
with 
\begin{align}
\mathcal{Z}_{\a\ad}=&~6C^{\g(2)\d(2)}h_{\d(2)\a\ad\bd(2)}\nabla_{\g}{}^{\bd}\nabla_{\g}{}^{\bd}\lambda+4\lambda C_{\a}{}^{\g(3)}\bar{C}_{\ad}{}^{\gd(3)}h_{\g(3)\gd(3)}\notag\\
&+6\nabla_{\g}{}^{\bd}\lambda\nabla^{\b\bd}\big[C_{\a}{}^{\g(3)}h_{\g(2)\b\bd(2)\ad}\big]-6\lambda\nabla_{\g}{}^{\bd}\nabla^{\b\bd}\big[C_{\a}{}^{\g(3)}h_{\g(2)\b\bd(2)\ad}\big]~.
\end{align}
Typically, the action that we begin with is primary (as it is here), which means that all conformal covariant derivatives in the action take the form
\begin{align}
\nabla_{\a\ad}=\mathcal{D}_{\a\ad}-\frac{1}{4}P_{\a\ad,}{}^{\b\bd}K_{\b\bd}
\end{align}
where $\mathcal{D}_{\a\ad}$ is the torsion-free Lorentz covariant derivative and $P_{\a\ad,\b\bd}$ is the Schouten tensor. Since we can always ignore total derivatives from the former, this allows us to rewrite \eqref{A.2} as 
\begin{align}
\mathcal{I}_{\text{Total}}=-\frac{1}{12}\int\text{d}^4x \, e \,P^{\a\ad,\b\bd}K_{\b\bd}\mathcal{Z}_{\a\ad}~.\label{A.5}
\end{align}
The above expression vanishes in most cases because $\mathcal{Z}_{\a\ad}$ turns out to be primary, however this is not true for the current example and one can instead show that \eqref{A.5} reduces to
\begin{align}
\mathcal{I}_{\text{Total}}=2\int\text{d}^4x \, e \,P^{\a\ad,\b\bd}\nabla_{\a}{}^{\gd}\bigg\{\lambda C_{\b}{}^{\g(3)}h_{\g(3)\ad\bd\gd}\bigg\}~. \label{A.6}
\end{align}
By making use of the well-known Bianchi identity
\begin{align}
\mathcal{D}^dC_{abcd}=-2\mathcal{D}_{[a}P_{b]c}\qquad \Longleftrightarrow \qquad \mathcal{D}_{\a}{}^{\bd}\bar{C}_{\ad(3)\bd}=\mathcal{D}_{(\ad_1}{}^{\b}P_{\b\ad_2,\a\ad_3)}~,
\end{align}
one can show that \eqref{A.6} is equivalent to
\begin{align}
\mathcal{I}_{\text{Total}}=2\int\text{d}^4x \, e \, \lambda\bigg\{C^{\a(3)\b}\nabla_{\b\bd}\bar{C}^{\ad(3)\bd}h_{\a(3)\ad(3)}\bigg\}~.
\end{align}
One must be careful to include this term when computing the gauge variation \eqref{2.22}. This subtlety regarding integration by parts does not occur elsewhere throughout this paper. 

\begin{footnotesize}

\end{footnotesize}


\begin{thebibliography}{66}

\bibitem{FT} 
E.~S.~Fradkin and A.~A.~Tseytlin,  ``Conformal supergravity,''
Phys.\ Rept.\  {\bf 119}, 233 (1985).

\bibitem{HST}
P.~S.~Howe, K.~S.~Stelle and P.~K.~Townsend,
``Supercurrents,''  Nucl.\ Phys.\  B {\bf 192}, 332 (1981).

\bibitem{KMT} 
S.~M.~Kuzenko, R.~Manvelyan and S.~Theisen,
``Off-shell superconformal higher spin multiplets in four dimensions,''
JHEP {\bf 1707}, 034 (2017)
  [arXiv:1701.00682 [hep-th]].


\bibitem{FL-algebras} 
  E.~S.~Fradkin and V.~Y.~Linetsky,
  ``Conformal superalgebras of higher spins,''
  Mod.\ Phys.\ Lett.\ A {\bf 4}, no. 24, 2363 (1989);
  ``Conformal superalgebras of higher spins,''
  Annals Phys.\  {\bf 198}, 252 (1990).

\bibitem{FL-vertices} 
E.~S.~Fradkin and V.~Y.~Linetsky,
``Cubic interaction in conformal theory of integer higher-spin fields in 
four dimensional space-time,''  Phys.\ Lett.\ B {\bf 231}, 97 (1989);
``Superconformal higher spin theory in the cubic approximation,''
  Nucl.\ Phys.\ B {\bf 350}, 274 (1991).

\bibitem{FV1} 
  E.~S.~Fradkin and M.~A.~Vasiliev,
  ``Candidate to the role of higher spin symmetry,''
  Annals Phys.\  {\bf 177}, 63 (1987).

\bibitem{FV2} 
  E.~S.~Fradkin and M.~A.~Vasiliev,
  ``Superalgebra of higher spins and auxiliary fields,''
  Int.\ J.\ Mod.\ Phys.\ A {\bf 3}, 2983 (1988).

\bibitem{Vasiliev88} 
  M.~A.~Vasiliev,
  ``Extended higher spin superalgebras and their realizations in terms of quantum operators,''
  Fortsch.\ Phys.\  {\bf 36}, 33 (1988).

\bibitem{FV-vertices} 
  E.~S.~Fradkin and M.~A.~Vasiliev,
  ``On the gravitational interaction of massless higher spin fields,''
  Phys.\ Lett.\ B {\bf 189}, 89 (1987);
%
  ``Cubic interaction in extended theories of massless higher spin fields,''
  Nucl.\ Phys.\ B {\bf 291}, 141 (1987).



\bibitem{Tseytlin} 
  A.~A.~Tseytlin,
  ``On limits of superstring in AdS(5) x S**5,''
  Theor.\ Math.\ Phys.\  {\bf 133}, 1376 (2002)
  [Teor.\ Mat.\ Fiz.\  {\bf 133}, 69 (2002)]
  [hep-th/0201112].


 \bibitem{Segal} 
  A.~Y.~Segal,
  ``Conformal higher spin theory,''
  Nucl.\ Phys.\ B {\bf 664}, 59 (2003)
  [hep-th/0207212]. 


\bibitem{BJM1} 
  X.~Bekaert, E.~Joung and J.~Mourad,
  ``On higher spin interactions with matter,''
  JHEP {\bf 0905}, 126 (2009)
  [arXiv:0903.3338 [hep-th]].


\bibitem{BJM2} 
  X.~Bekaert, E.~Joung and J.~Mourad,
  ``Effective action in a higher-spin background,''
  JHEP {\bf 1102}, 048 (2011)
  [arXiv:1012.2103 [hep-th]].

\bibitem{Bonezzi} 
  R.~Bonezzi,
  ``Induced action for conformal higher spins from worldline path integrals,''
  Universe {\bf 3}, no. 3, 64 (2017)
  [arXiv:1709.00850 [hep-th]].

\bibitem{KP2} 
  S.~M.~Kuzenko and M.~Ponds,
  ``Conformal geometry and (super)conformal higher-spin gauge theories,''
  JHEP {\bf 1905}, 113 (2019)
  [arXiv:1902.08010 [hep-th]].
  
 
\bibitem{DeserN} 
S.~Deser and R.~I.~Nepomechie,
   ``Gauge invariance versus masslessness in de Sitter space,''
  Annals Phys.\  {\bf 154}, 396 (1984).


\bibitem{DeserN2}
  S.~Deser and R.~I.~Nepomechie,
  ``Anomalous propagation of gauge fields in conformally flat spaces,''
  Phys.\ Lett.\  {\bf 132B} (1983) 321.

\bibitem{Vasiliev2006}
  E.~D.~Skvortsov and M.~A.~Vasiliev,
  ``Geometric formulation for partially massless fields,''
  Nucl.\ Phys.\ B {\bf 756} (2006) 117
  [hep-th/0601095].
  
\bibitem{Vasiliev2009} 
  M.~A.~Vasiliev,
  ``Bosonic conformal higher-spin fields of any symmetry,''
  Nucl.\ Phys.\ B {\bf 829}, 176 (2010)
  [arXiv:0909.5226 [hep-th]].

\bibitem{DNW} 
  L.~Dolan, C.~R.~Nappi and E.~Witten,
  ``Conformal operators for partially massless states,''
  JHEP {\bf 0110}, 016 (2001)
  [hep-th/0109096]. 

\bibitem{DeserW5} 
  S.~Deser and A.~Waldron,
  ``Conformal invariance of partially massless higher spins,''
  Phys.\ Lett.\ B {\bf 603}, 30 (2004)
  [hep-th/0408155].
  
\bibitem{DeserW6} 
  S.~Deser, E.~Joung and A.~Waldron,
  ``Partial masslessness and conformal gravity,''
  J.\ Phys.\ A {\bf 46}, 214019 (2013)
  [arXiv:1208.1307 [hep-th]].  


\bibitem{BG2013} 
  X.~Bekaert and M.~Grigoriev,
  ``Higher order singletons, partially massless fields and their boundary values in the ambient approach,''
  Nucl.\ Phys.\ B {\bf 876}, 667 (2013)
  [arXiv:1305.0162 [hep-th]]. 
  
\bibitem{GrigorievH} 
  M.~Grigoriev and A.~Hancharuk,
  ``On the structure of the conformal higher-spin wave operators,''
  JHEP {\bf 1812}, 033 (2018)
  [arXiv:1808.04320 [hep-th]].
 
   \bibitem{Skvortsov}
  M.~Grigoriev, I.~Lovrekovic and E.~Skvortsov,
  ``New conformal higher spin gravities in $3d$,''
  [arXiv:1909.13305 [hep-th]].

\bibitem{KTvN1} 
  M.~Kaku, P.~K.~Townsend and P.~van Nieuwenhuizen,
  ``Gauge theory of the conformal and superconformal group,''
  Phys.\ Lett.\  {\bf 69B}, 304 (1977).

 
\bibitem{NT} 
  T.~Nutma and M.~Taronna,
  ``On conformal higher spin wave operators,''
  JHEP {\bf 1406}, 066 (2014)
  [arXiv:1404.7452 [hep-th]].  
  

\bibitem{GrigorievT} 
  M.~Grigoriev and A.~A.~Tseytlin,
  ``On conformal higher spins in curved background,''
  J.\ Phys.\ A {\bf 50}, no. 12, 125401 (2017)
  [arXiv:1609.09381 [hep-th]].  
  
\bibitem{BeccariaT} 
  M.~Beccaria and A.~A.~Tseytlin,
  ``On induced action for conformal higher spins in curved background,''
  Nucl.\ Phys.\ B {\bf 919}, 359 (2017)
  [arXiv:1702.00222 [hep-th]].  
  
\bibitem{Manvelyan} 
R.~Manvelyan and G.~Poghosyan,
``Geometrical structure of Weyl invariants for spin three gauge field in general gravitational background in $d=4$,'' Nucl.\ Phys.\ B {\bf 937}, 1 (2018)
  [arXiv:1804.10779 [hep-th]].  


\bibitem{BKNT-M1} 
D.~Butter, S.~M.~Kuzenko, J.~Novak and G.~Tartaglino-Mazzucchelli,
``Conformal supergravity in three dimensions: New off-shell formulation,''
JHEP {\bf 1309}, 072 (2013)
  [arXiv:1305.3132 [hep-th]].
  
  \bibitem{BK} I.~L.~Buchbinder and S.~M.~Kuzenko,
{\it Ideas and Methods of Supersymmetry and
Supergravity or a Walk Through Superspace}, IOP, Bristol, 1995
(Revised Edition: 1998).

\bibitem{BT2015} 
M.~Beccaria and A.~A.~Tseytlin, ``On higher spin partition functions,''
  J.\ Phys.\ A {\bf 48}, no. 27, 275401 (2015)
  [arXiv:1503.08143 [hep-th]].  


\bibitem{LN} 
  A.~A.~Leonovich and V.~V.~Nesterenko,
  ``Conformally invariant equation for the symmetric tensor field,''
Dubna prieprint  JINR-E2-84-11.

 \bibitem{Sachs} 
 A.~Iorio, L.~O'Raifeartaigh, I.~Sachs and C.~Wiesendanger,
``Weyl gauging and conformal invariance,''
  Nucl.\ Phys.\ B {\bf 495}, 433 (1997)
  [hep-th/9607110].

 \bibitem{EO} 
  J.~Erdmenger and H.~Osborn,
  ``Conformally covariant differential operators: Symmetric tensor fields,''
  Class.\ Quant.\ Grav.\  {\bf 15}, 273 (1998)
  [gr-qc/9708040].
  
 \bibitem{KPR} 
  S.~M.~Kuzenko, M.~Ponds and E.~S.~N.~Raptakis, ``New locally (super)conformal gauge models
in Bach-flat backgrounds,'' [arXiv:2005.08657 [hep-th]].



\end{thebibliography}
\end{document}